# Massive Dirac fermions and Hofstadter butterfly in a van der Waals heterostructure

B. Hunt,[1*] J. D. Sanchez-Yamagishi,[1*] A. F. Young,[1*]
K. Watanabe,[2] T. Taniguchi,[2] P. Moon[3], M. Koshino[3],
P. Jarillo-Herrero[1†] and R. C. Ashoori.[1†]

[1]Department of Physics, Massachusetts Institute of Technology Cambridge, MA, USA
[2]Advanced Materials Laboratory, National Institute for Materials Science, Tsukuba, Japan
[3]Department of Physics, Tohoku University, Sendai, Japan
*These authors contributed equally to the work. †pjarillo@mit.edu; ashoori@mit.edu

Van der Waals heterostructures comprise a new class of artificial materials formed by stacking atomically-thin planar crystals. Here, we demonstrate band structure engineering of a van der Waals heterostructure composed of a monolayer graphene flake coupled to a rotationally-aligned hexagonal boron nitride substrate. The spatially-varying interlayer atomic registry results both in a local breaking of the carbon sublattice symmetry and a long-range moiré superlattice potential in the graphene. This interplay between short- and long-wavelength effects results in a band structure described by isolated superlattice minibands and an unexpectedly large band gap at charge neutrality, both of which can be tuned by varying the interlayer alignment. Magnetocapacitance measurements reveal previously unobserved fractional quantum Hall states reflecting the massive Dirac dispersion that results from broken sublattice symmetry. At ultra-high fields, integer conductance plateaus are observed at non-integer filling factors due to the emergence of the Hofstadter butterfly in a symmetry-broken Landau level.

The ability to tailor the properties of electronic devices is one of the landmark achievements of modern technology, and underlies much of modern research in condensed matter physics. Just as crystal structure can determine the electronic properties of a material, artificial periodic superstructures can be used to modify the electronic band structure of existing materials [1]. The band structure of pristine graphene consists of linearly dispersing energy bands, which touch at two degenerate "Dirac points". This degeneracy is protected by the equivalence of the A and B triangular sublattices that make up the graphene honeycomb [2], and is responsible for graphene's semimetallic behavior. Theory suggests that the electronic properties of graphene can be tuned via external periodic potentials: long-wavelength superlattices have been predicted to lead to the formation of additional gapless Dirac points at finite energy[3], while atomic scale modulations, by breaking the A-B sublattice symmetry, may turn graphene from a semimetal into a semiconductor[4]. Experimental efforts to make high mobility functional devices based on band structure engineering, however, have been hindered by growth and nanofabrication limitations[5].

Recently, a new approach has become available through the use of hexagonal boron nitride (hBN) as a planar crystalline substrate. hBN is isostructural to graphene, but has boron and nitrogen atoms on the A and B sublattices leading to a band gap in the electronic structure[6]. The weak interlayer van der Waals forces in both graphene and hBN permit the fabrication of multilayer heterostructures by sequential transfer of individual layers[7]. During the transfer process, the angular alignment of the constituent crystals ($\theta$) can in principle be controlled, but the graphene and hBN lattices retain their natural 1.8% mismatch [4]. The beating of the mismatched lattices leads to the formation of a moiré pattern with wavelength $\lambda(\theta)$ that can be much larger than the lattice constant [8](See Fig. 1A and [9]).

The effect of the moiré on the graphene electronic structure can be decomposed into two parts[10]. The moiré pattern results in a $\lambda$-scale modulation of the graphene-hBN coupling, forming a smooth superlattice potential. More subtly, the moiré also modulates the local asymmetry between the graphene sublattices induced by the difference in potential between boron and nitrogen atoms in the hBN. The resulting A-B potential difference in the graphene, parameterized in Fig 1A as $m(\vec{r})$, oscillates across the superlattice unit cell, leading to nearly complete cancellation[10] upon spatial average. As we demonstrate, however, the absence of sublattice symmetry nonetheless has dramatic experimental consequences for the electronic properties near the charge neutrality point.

We present measurements of four heterostructure devices consisting of a monolayer graphene flake on a 7 nm-thick hBN substrate[9], which itself sits on top of a graphite local gate (Fig 1B). The proximal gate electrode serve both as an extremely flat substrate and to screen long-range potential fluctuations in the graphene[11], leading to high quality devices with field effect mobilities of $\sim$100,000 cm$^2$/V·s and well-quantized quantum Hall plateaus at fields $B \lesssim$100 mT[9]. In contrast to the vast majority of graphene devices, which, with few exceptions[11, 12], are semimetallic with zero-field minimum conductivity $\sim 2e^2/h$, all four devices are strongly insulating near the overall charge neutrality point (CNP) (Fig. 1B). In addition, two devices show pronounced resistance peaks at finite density (Fig. 1C), situated symmetrically about the CNP. Notably, the devices showing additional resistance peaks also have the strongest insulating states.

We ascribe the satellite resistance peaks to the Bragg scattering of charge carriers by the superlattice when the lowest electron and hole minibands are fully occupied [8, 13–15]. In graphene, due to the spin and valley degeneracies, full filling occurs at a density of four electrons per superlattice unit cell, $n = 4n_0$, where $1/n_0 = \sqrt{3}\lambda^2/2$ is the unit cell area (see Fig.

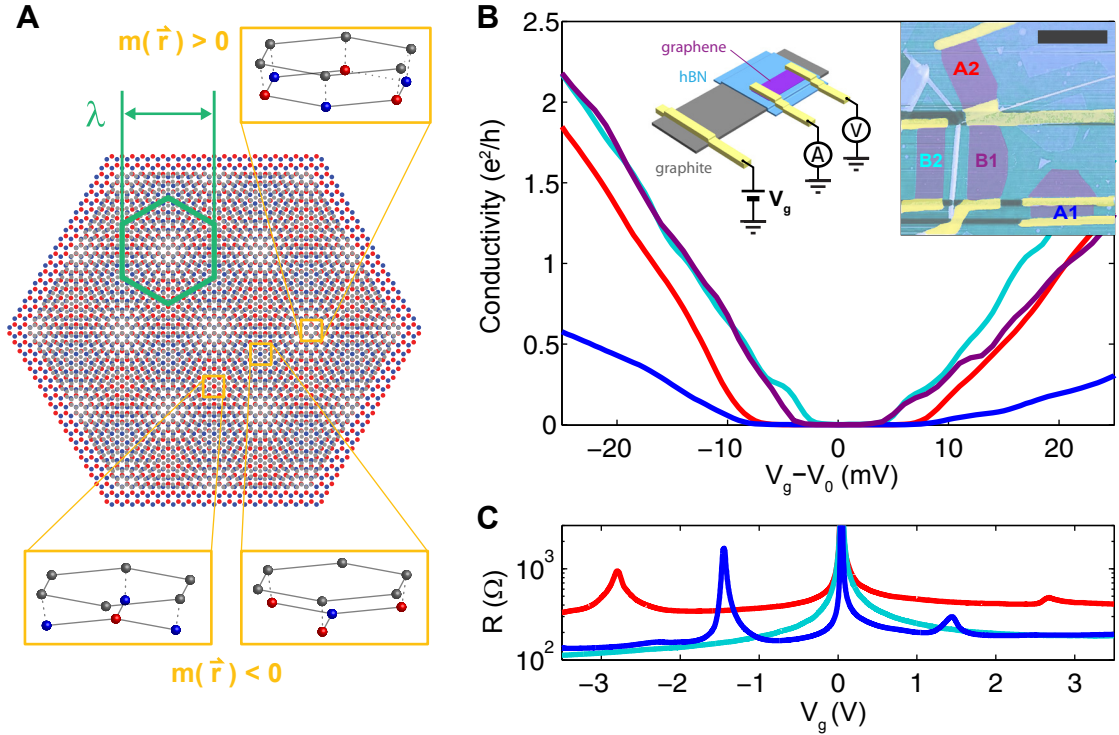

Figure 1. **Insulating states and superlattice minibands in a graphene/hBN heterostructure** (A) Schematic of the moiré pattern for graphene (gray) on hBN (red and blue), for zero misalignment angle and an exaggerated lattice mismatch of $\sim 10\%$. The moiré unit cell is outlined in green. Regions of local quasi-epitaxial alignment lead to opposite signs of the sublattice asymmetry, $m(\vec{r})$, in different regions. (B) Low temperature ($T$=150 mK) conductivity near charge neutrality of four heterostructure devices. The CNP offset $V_0$=37, 37, 46 and 42 mV respectively for A1, A2, B1 and B2. Left inset: Measurement schematic. Right inset: AFM image. Scale bar is 3 $\mu$m. (C) Resistance over a larger gate range. Finite-density resistance peaks indicate full filling of superlattice minibands in two of the four measured devices (A1 and A2) within the experimentally-accessible density range.

1A). Using the density at which the peaks are observed we estimate $\lambda_{A1}$=11-13.5nm and $\lambda_{A2}$=7.5-9.5nm, where the error is dominated by uncertainty in the value of the hBN dielectric constant[9]. Such large values of $\lambda$, close to the theoretical maximum of 14nm, imply nearly perfect rotational alignment of the graphene and hBN, with $\theta < 2°$ for the two devices showing satellite peaks. Given that all four devices are fabricated from a single transferred graphene flake, we expect their orientations to be correlated, suggesting that devices B1 and B2 are also closely aligned with the substrate even though the putative superlattice miniband edge falls outside the observable range.

The moiré pattern offers a unique opportunity to study the elementary problem of a charged quantum particle moving under the simultaneous influence of a periodic potential and a magnetic field[16–19] in the normally inaccessible regime of more than one magnetic flux quantum ($\phi_0$) per superlattice unit cell. In the absence of the superlattice, graphene is described at high fields by a set of discrete, highly degenerate Landau levels (LLs) indexed by an integer $N$. The periodic potential splits the flat LL bands into "Hofstadter minibands", separated by a hierarchy of self-similar minigaps[17].

Despite the intricate structure of the Hofstadter spectrum, the densities corresponding to the fractal minigaps follow simple linear trajectories as a function of magnetic field[20]. Magnetoresistance data indeed show strong effects of the superlattice (Fig. 2A), including Landau fans originating from both the central and satellite zero-field resistance peaks. As recently demonstrated, the intersections between the central and satellite fans occur at $\phi = \phi_0/q$ (Fig. 2B), where $\phi$ is the magnetic flux per superlattice unit cell and $q$ is a positive integer [14, 15]. These intersections allow a second, independent method of measuring the unit cell area without reference to electrostatic parameters[9], giving $\lambda_{A1} = 12.9 \pm 0.2$ nm and $\lambda_{A2} = 9.2 \pm 0.1$ nm.

The full development of the Hofstadter butterfly, however, is most obvious in the regime $\phi/\phi_0 > 1$, which has not previously been accessed in monolayer graphene. Figure 2C shows the conductance within the $N$=0 Landau level for two values of field corresponding to $\phi < \phi_0$ and $\phi > \phi_0$. At the higher field, the $N$=0 LL is completely reconstructed, with a nonmonotonic sequence of conductance plateaus that is well matched by tight-binding calculations[9] of the Hofstadter butterfly spectrum in which phenomenological spin-and sublattice-breaking terms have been included (Fig. 2D). The emergence of states with integer quantized conductance at noninteger filling of a single Landau level, severing the canonical relationship between quantized conductance and filling fraction, is the signature of the Hofstadter butterfly.

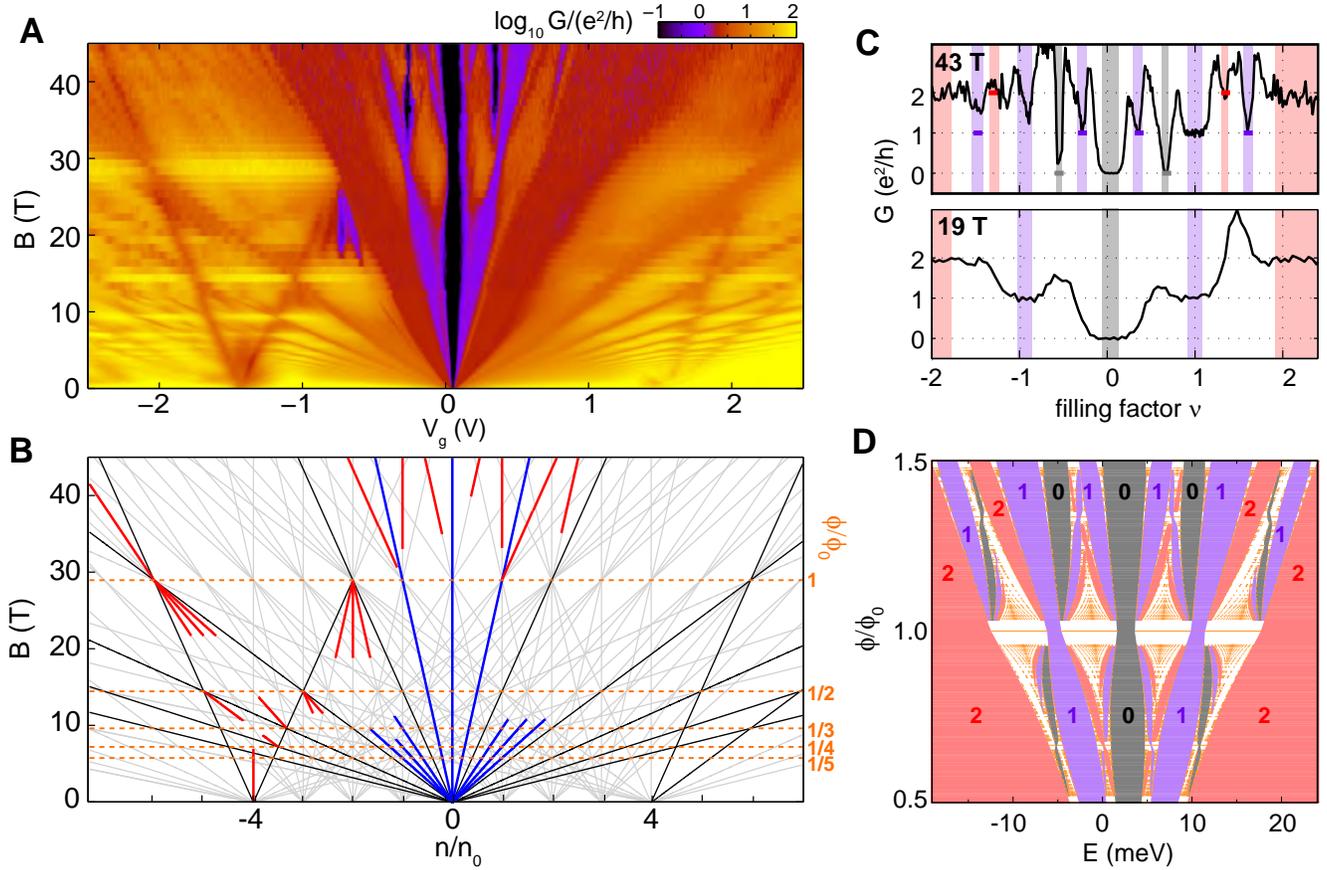

Figure 2. **Hofstadter butterfly.** (**A**) Two-terminal magnetoconductance of device A1 up to 45 T. (**B**) Energy gaps in the Hofstadter spectrum are confined to linear trajectories $\phi/\phi_0 = (n/n_0 - s)/t$, where $s$ and $t$ are integers denoting the superlattice miniband filling index [32] and quantized Hall conductance of the gapped state, respectively. Grey lines indicate gaps for $-4 \leq s \leq 4$, with colored overlays indicating features observed in (A). Black: gaps requiring no broken symmetry; Blue: symmetry-broken states for the central Landau fan. Red: symmetry broken states belonging to superlattice ($s \neq 0$) Landau fans. Gaps intersect at $\phi/\phi_0 = 1/q$, with $q$ an integer (orange); $\phi = \phi_0$ at 29T. (**C**) Conductance traces within the $N = 0$ LL at B=43T (top) and B=19 T(bottom), showing the emergence of Hofstadter minigaps for $\phi > \phi_0$. Shaded rectangles are color coded to the two terminal conductance expected from the Hofstadter model. Peaks between plateaus are due to diffusive transport in this wide aspect ratio device. (**D**) Theoretical Hofstadter energy spectrum based on fully spin- and sublattice- split Landau levels. Orange points indicate regions of dense energy bands and spectral gaps are color coded to the two-terminal conductance[9, 33].

Equating the effect of the hBN substrate with that of a smooth superlattice potential explains many features of the experimental data, including the satellite resistance peaks and most features of the ultra-high $B$ transport data. However, it fails to account for the insulating state observed at charge neutrality, which persists uninterrupted from $B$=0 to 45 T. To further explore the properties of this state, we measure the capacitance of the graphene to the proximal graphite backgate using a low-temperature capacitance bridge[9, 21]. Capacitance measurements probe the thermodynamic density of states, $\partial n/\partial \mu$; for our parallel plate geometry, the measured capacitance $C_{meas}^{-1} = C_{geom}^{-1} + (Ae^2 \partial n/\partial \mu)^{-1}$, where $C_{geom}$ is the geometric capacitance and $A$ is the sample area [22]. Figure 3A shows magnetocapacitance data from a typical, semimetallic graphene-on-hBN device. The capacitance, and by extension the density of states, has a minimum at charge neutrality. As the field is increased, this minimum is replaced by a local maximum, signifying the formation of the zero-energy Landau level characteristic of massless Dirac fermions[23]. Capacitance measurements of an insulating graphene device reveal very different behavior (Fig. 3B). No peak forms at the CNP at finite field, as can be seen from the dark vertical region centered at $V_g = 41$ mV, indicating that the $N$=0 Landau is split into two finite-energy sublevels; in other words, a Landau level never forms at zero energy.

As we increase the field further (Fig. 3C), additional minima in capacitance appear at all integer filling factors, $\nu$, including those not belonging to the standard monolayer graphene sequence, which indicates the emergence of exchange-driven broken symmetry states [24]. Capacitance minima associated with fractional quantum Hall states appear (Fig. 3D) at all multiples of 1/3 for -2< $\nu$ <2. Notably, this sequence includes robust features at $\nu = \pm\frac{5}{3}$ not observed in previous studies [25, 26] of semimetallic monolayer graphene. These states are thought to be described by a fully spin- and sublattice-polarized Laughlin wavefunction [27, 28]; their absence in semimetallic graphene is attributed[25, 29] to the low energy cost of exciting charge carriers to the unoccupied, energetically equivalent sublattice. Our observation of the $\pm\frac{5}{3}$ states suggests that sublattice symmetry is broken in our graphene-hBN heterostructures.

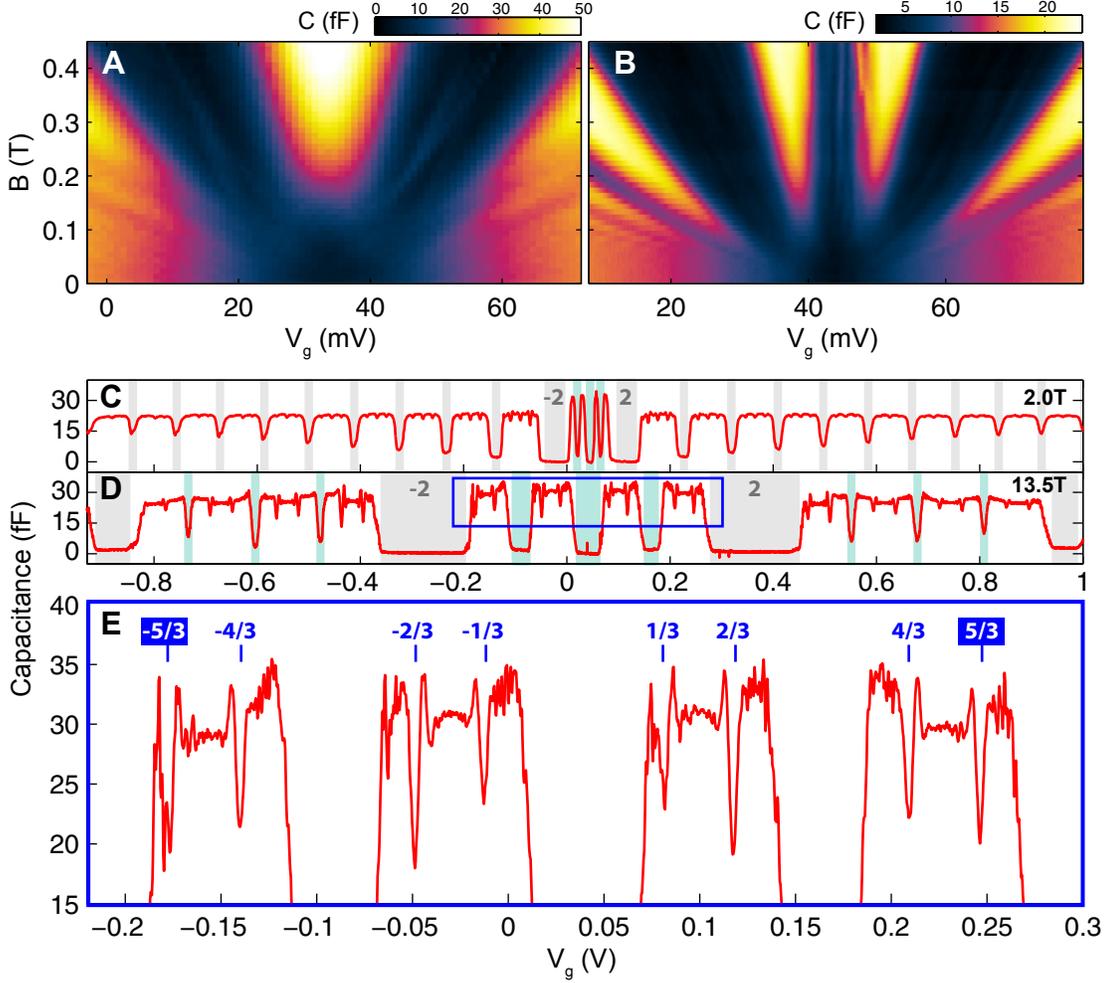

Figure 3. **Magnetocapacitance of semimetallic and insulating graphene devices.** (A) Capacitance of a typical semimetallic graphene device on an 8.4 nm-thick hBN flake. The zero-energy Landau level forms at $\sim 0.15$T, appearing as a local maximum at the CNP ($V_g = 34.5$ mV). (B) Capacitance of an insulating graphene device (B2) under similar conditions. In contrast to (A), the density of states is always at a local minimum at charge neutrality ($V_g = 44$ mV). (C) Capacitance of device B2 at $B$=2.0 T and (D) 13.5 T. Cyclotron gaps are shaded grey ($\nu = \pm 2$ labeled) and broken-symmetry gaps are shaded green. (E) Fractional quantum Hall (FQH) states in the $N = 0$ LL. The incompressible features at $\nu = \pm\frac{5}{3}$ are of similar strength to the other FQH states.

In the Dirac equation description of graphene, sublattice symmetry breaking can be parameterized by a mass. In our heterostructures, this mass term $m(\vec{r})$ is expected to oscillate across the moiré unit cell (Fig. 1A). Remarkably, the low density phenomenology of our insulating graphene at low fields ($\phi \ll \phi_0$), including the insulating gap, the absence of a zero-energy Landau level, and the observation of the $\nu = \pm\frac{5}{3}$ states, can be captured by a Dirac equation with a spatially-uniform *global* effective mass, $m^*$. The resulting Hamiltonian describing physics in the vicinity of the $K(K')$ point is

$$\hat{H} = v_F \boldsymbol{\sigma} \cdot \boldsymbol{p} + m^* v_F^2 \hat{\sigma}_z \quad (1)$$

$$= \begin{pmatrix} \pm m^* v_F^2 & v_F(p_x - ip_y) \\ v_F(p_x + ip_y) & \mp m^* v_F^2 \end{pmatrix} \quad (2)$$

where $v_F$ is the Fermi velocity, $\boldsymbol{\sigma} = (\sigma_x, \sigma_y)$, and the Pauli matrices $\{\sigma_i\}$ operate in the basis of the two sublattices. The resulting energy spectrum at zero magnetic field, $E(p) = \pm\sqrt{v_F^2 p^2 + (m^* v_F^2)^2}$, features a band gap $\Delta = 2m^*v_F^2$ at charge neutrality. In a quantizing magnetic field, the Landau level spectrum is given by $E_N = \pm\sqrt{2(\hbar v_F)^2 |N|/\ell_B^2 + (m^*v_F^2)^2}$, where $\ell_B = \sqrt{\hbar/(eB)}$ is the magnetic length [9]. The mass term does not lift the LL degeneracy for $|N|>0$; this is reflected in our data (Fig. 3C) by the observation of symmetry breaking in the higher Landau levels only at higher magnetic fields, presumably due to exchange interactions, as in semimetallic graphene[24]. However, the splitting of the $N=0$ LL into two sublattice-polarized branches at $E_0 = \pm m^*v_F^2$ is consistent with the persistent gap at charge neutrality, as well as the observation of the $\pm\frac{5}{3}$ fractional quantum Hall states.

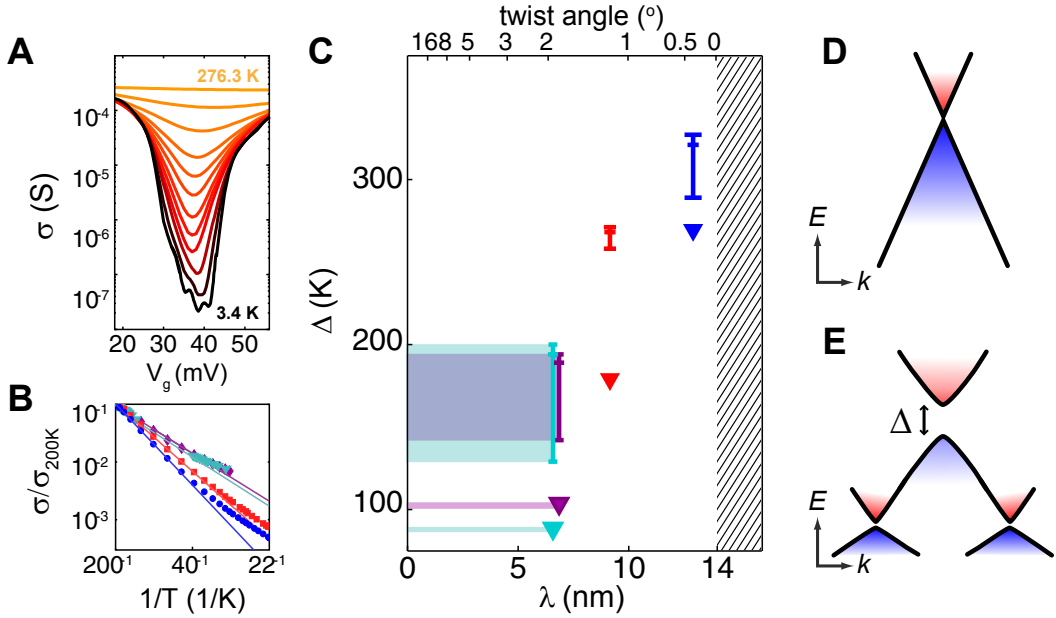

Figure 4. **Energy gaps of the zero-field insulator.** (A) Temperature dependence of the insulator (device A1). (B) Arrhenius plot of $\sigma_{CNP}$ for all four devices. Lines are fits to $\sigma_{CNP}(T) \propto \exp(-\Delta/2T)$, giving $\Delta_{A1}$=320K, $\Delta_{A2}$=270K, $\Delta_{B1}$=190K, and $\Delta_{B2}$=195K. (C) Correlation of observed band gaps with moiré wavelength $\lambda$. Dashes: thermal activation gap (Fig. 2B). Triangles: upper bound on chemical potential change $\Delta\mu$[9]. $\lambda$ is extracted from the period of the Hofstadter oscillations. For devices B1 and B2, the gate-leakage limited maximum density sets an upper bound of $\lambda \simeq 6.5$ nm. Cross-hatching: theoretically inaccessible region set by the graphene-hBN lattice mismatch. (D) Schematic band structure for semimetallic graphene. (E) Schematic band structure for a graphene-hBN heterostructure with a small twist angle (not to scale), showing the band gap and moiré minibands.

The observation of band gaps in samples with long wavelength moiré patterns is not likely to be coincidental. Naively, $m(\vec{r})$ nearly vanishes upon spatial average, calling into question whether a mismatched hBN substrate can open a measurable gap. Indeed, predictions for how the global parameter $m^*$ depends on the microscopic structure of $m(\vec{r})$ vary by several orders of magnitude depending on theoretical assumptions[4, 10, 30, 31]. We quantitatively assess the connection between $\lambda$ and $m^*$ by measuring the band gap using thermally-activated transport (Fig. 4A-B) and, independently, by measuring the width in gate voltage of the insulating state, which places an upper bound on the chemical potential difference across the band gap[9]. Figure 4C shows a correlation between moiré wavelength and the measured gaps, suggesting that the interaction of the graphene flake with the closely-aligned hBN substrate is responsible for the insulating behavior. The discrepancy between the two methods for quantifying the gap, which are not equally sensitive to the role of Fermi-energy dependent electronic interactions, may lend support to the recent proposal[31] that the huge enhancements of the measured band gap relative to the predictions of noninteracting theory[10] may have a many-body origin.

The ability to engineer a nonzero band mass in graphene has far-reaching implications for future experimental efforts. The possibility of an alignment-dependent mass may require a reinterpretation of experiments involving graphene-hBN heterostructures, even as it engenders new opportunities for the design of electronic devices. The gapped spectrum provides a useful tool in nanoengineering based on electrostatic confinement, while the tunability of the effective mass provides both a new probe and design parameter in the study of many-body effects. Finally, the possible many-body enhancements of the measured band gaps in our mismatched graphene-hBN bilayer raises the tantalizing possibility that epitaxial graphene-hBN structures, if susceptible to similar effects, might give rise to even larger, technologically-relevant band gaps.

*Acknowledgements.* We acknowledge helpful discussions with C. Dean, H. Churchill, L. Levitov, and J. Song. B.H. and R.C.A. were funded by the BES Program of the Office of Science of the US DOE, Contract No. FG02-08ER46514 and the Gordon and Betty Moore Foundation through Grant GBMF2931. J.D.S-Y, and P.J-H. have been primarily supported by the US DOE, BES Office, Division of Materials Sciences and Engineering under Award DE-SC0001819. Early fabrication feasibility studies were supported by NSF Career Award No. DMR-0845287 and the ONR GATE MURI. This work made use of the MRSEC Shared Experimental Facilities supported by NSF under award No. DMR-0819762 and of Harvard's CNS, supported by NSF under grant No. ECS-0335765. Some measurements were performed at the National High Magnetic Field Laboratory, which is supported by NSF Cooperative Agreement DMR-0654118, the State of Florida and DOE.

# Supplementary Information:
# Massive Dirac fermions and Hofstadter butterfly in a van der Waals heterostructure

B. Hunt,[1*] J. D. Sanchez-Yamagishi,[1*] A. F. Young,[1*]
K. Watanabe,[2] T. Taniguchi,[2] P. Moon,[3] M. Koshino,[3],
P. Jarillo-Herrero[1†] and R. C. Ashoori.[1†]

[1]Department of Physics, Massachusetts Institute of Technology Cambridge, MA, USA
[2]Advanced Materials Laboratory, National Institute for Materials Science, Tsukuba, Japan
[3]Department of Physics, Tohoku University, Sendai, Japan
*These authors contributed equally to the work.  †pjarillo@mit.edu; ashoori@mit.edu

## CONTENTS

## I. ZERO AND LOW FIELD TRANSPORT MEASUREMENTS

Figure S1 shows magnetotransport measurements performed on the four semiconducting graphene devices at the base temperature of 150 mK. They attest to the high quality of our devices: well-formed plateaus at filling factor $\nu = 2$ appear for $B_Q \approx 100$mT, implying that the quantum mobility $\mu_Q = 1/B_Q$ is at least $10^5$ cm$^2$/(V·s). This is in agreement with our measurements of the field-effect mobility $\mu_{FE} = e^{-1}(d\sigma/dn)$ close to charge neutrality (Fig S2). Note that the field-effect mobility for all devices is an underestimate due to the effects of contact resistance and quantum capacitance, which reduces the capacitance used in calculating the charge density $n$ in this simple estimate.

The transport data in Figure S1A is the analogue of the capacitance data presented in the main text (Figure 3), and similarly shows the insulating state at zero field persisting as the field is raised. The insulator appears as a dark vertical band centered at the charge neutrality point (CNP), further confirming that a Landau level never forms at zero energy. A line trace of conductance at a single fixed density inside the gap (Fig. S1C) shows that the conductance is a monotonically decreasing function of $|B|$, decreasing by two orders of magnitude between $B = 0$ and $B = 2$ T before dropping below the noise floor of our measurement.

## II. MOIRÉ SUPERLATTICES

A graphene-hBN heterostructure naturally leads to a moiré superlattice due to the $\delta = 1.8\%$ lattice mismatch and the rotational misalignment between the layers (quantified by the twist angle $\theta$). The wavelength $\lambda$ of the moiré is related to $\theta$ by[1]

$$\lambda = \frac{(1+\delta)a}{\sqrt{2(1+\delta)(1-\cos\theta)+\delta^2}}, \qquad (1)$$

where $a$ is the graphene lattice constant.

The wavelength $\lambda$ of the moiré superlattice can be determined from its effects on the graphene magnetotransport measurements (Fig. S3 and S4). At zero field, $\lambda$ can be estimated from the location in gate voltage of the superlattice Dirac points, which occur at carrier density $n = 4n_0$, where $n_0$ is the inverse superlattice unit cell area. This requires knowledge of the carrier density as a function of gate voltage, which can be inferred from a charging model of the device that includes the effect of quantum capacitance $\tilde{C}_Q \equiv e^2(dn/d\mu)$ close to the CNP to give

$$n(V_g) = \frac{\tilde{C}_{geom}V_g}{e} - n_Q\left(\sqrt{1+\frac{\tilde{C}_{geom}V_g}{en_Q}} - 1\right), \qquad (2)$$

where $n_Q \equiv (\pi/2)(\tilde{C}_{geom}\hbar v_F/e^2)^2$ [2] and $\tilde{C} \equiv C/A$ denotes capacitance per unit area $A$ of the device. From the observed location of the superlattice Dirac points at 1.45V and 2.80V we estimate $\lambda$ =11-13.5nm and 7.5-9.5nm for samples A1 and A2, respectively. This simple model does not take into account any modifications of the graphene band structure by electron-electron interactions or the moiré, but these effects are small in comparison to the error due to the uncertainty in the hBN dielectric constant.

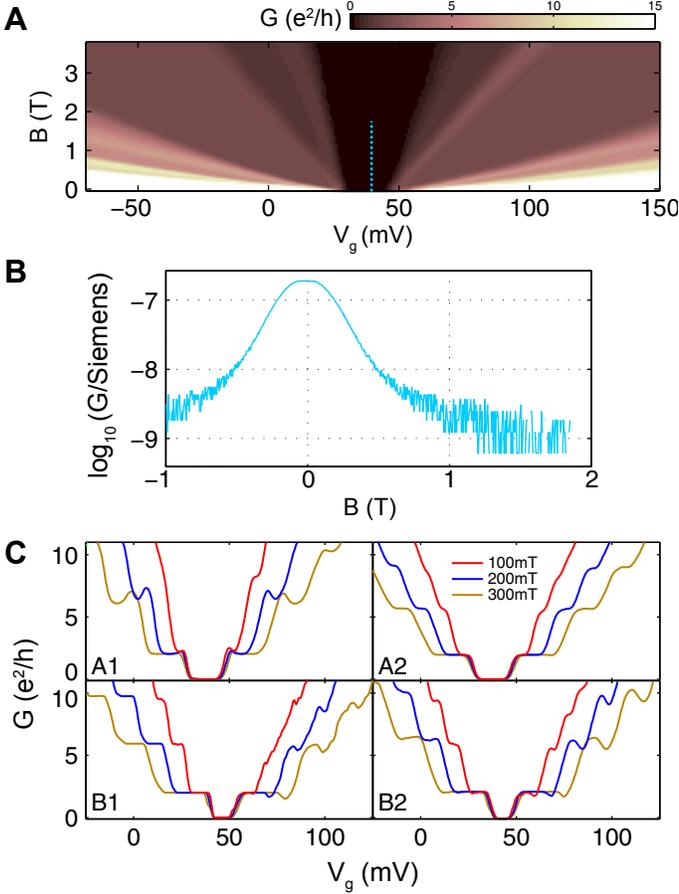

Figure S1. **Magnetotransport measurements of insulating graphene devices.** (A) Conductance of device A1 as a function of gate voltage and magnetic field. (B) Conductance trace at $V_g = 40$ mV, showing that $G_{CNP}$ is a monotonically decreasing function of $|B|$. (C) Gate sweeps at low field of the four devices, at $B$=100 (red), 200 (blue) and 300 mT (tan). Well-quantized $\nu = \pm 2$ plateaus appear at $B \lesssim 100$ mT for all devices.

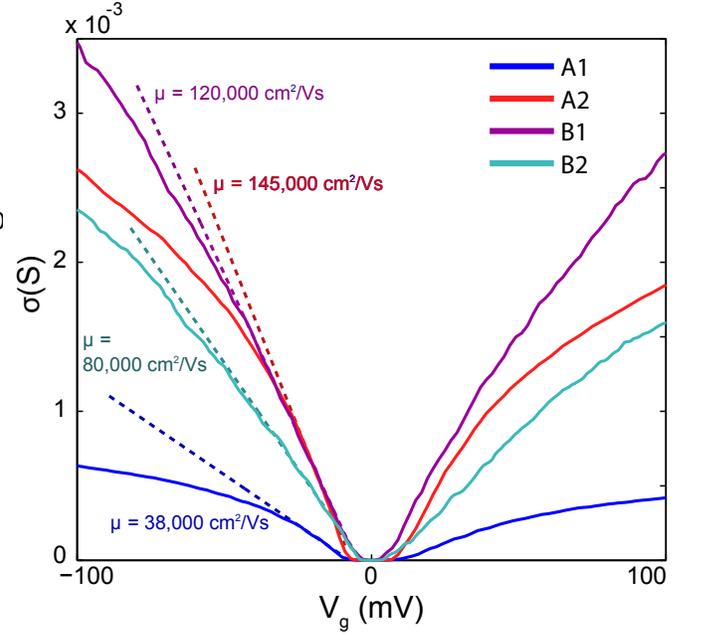

Figure S2. **Zero-field transport and field-effect mobility of insulating graphene devices** Conductivity vs. gate voltage for the four devices discussed in main text. Dashed lines are low density tangents whose slope corresponds to the field effect mobility

A more accurate method for extracting $\lambda$ is to fit the magnetotransport features associated with the Hofstadter spectrum. This method is purely geometric, and does not require knowledge of the carrier density, as it directly connects the superlattice unit cell area with the applied magnetic field. Figure S3 shows magnetotransport measurements of samples A1, A2, and B1. Data from both samples A1 and A2 show secondary Landau fans centered at the high density resistance peaks discussed in the main text, with a clear beating pattern arising from the interpenetrating Landau fans. When plotted on $V_G - 1/B$ axes (Fig. S4), it is clear that these bands of conductance peaks and intersecting gap features are uniformly spaced in $1/B$. Within the Wannier picture[3], integer Landau level gaps for different Landau fans are expected to intersect when $\phi/\phi_0 = 1/q$, where $q$ is an integer. For example, the $te^2/h$ gap features associated with the first Landau level of the $s = -4$ miniband and the $t'e^2/h$ state of the $s = 0$ fan will intersect at a density, such that

$$n/n_0 = t(\phi/\phi_0) - 4 = t'(\phi/\phi_0) \quad (3)$$
$$\phi/\phi_0 = -4/(t' - t) = 1/q \quad (4)$$

as long as $t$ and $t'$ are both integers in the main graphene sequence, $t^{(\prime)} = 4(N + 1/2)$. The positions of these features are fit by a single parameter $B_0$, the magnetic field at which one flux quantum threads the superlattice unit cell. Measuring $B_0$ allows us to infer the superlattice unit cell area $n_0^{-1}$ and by extension $\lambda = \sqrt{2/\sqrt{3}n_0}$. Fitting the data in Figure S4 produces values for samples A1 and A2 of $B_0 = 28.7 \pm 1.0$ T and $57.0 \pm 1.2$ T, giving $\lambda$=12.9 $\pm$ 0.2 nm and 9.2 $\pm$ 0.1 nm, respectively.

## III. THEORETICAL MODEL FOR HOFSTADTER SPECTRUM OF MONOLAYER GRAPHENE ON HBN

For the numerical calculation of the Hofstadter band structure shown in Fig. 2D of the main text, graphene and hBN are modeled by honeycomb lattices with lattice periods $a = 0.246$ nm and $a_{\mathrm{hBN}} = 0.2504$ nm, [6] respectively. We assume that graphene monolayer and hBN monolayer are aligned with zero rotation angle, and the ratio between the two lattice constants is round to a rational number $a_{\mathrm{hBN}}/a = 56/55$ to give terlayer distance between hBN and graphene is set to 0.322 nm [7]. We consider the $p$-type orbital state on each atomic site within the tight binding model, and set the on-site potential to 0, 3.34eV and $-1.40$eV for the C, B and N atoms, respectively[8]. For the hopping amplitudes between different sites, we adopt the Slater-Koster parametrization [9] irrespective of the atomic species under consideration. To compute the low-energy spectrum in magnetic field, we take the low-lying Landau levels ($|E| < 1.5$ eV) of isolated monolayer graphene as the basis [9], and the coupling with hBN states is included as an on-site potential on the graphene atomic sites within second-order perturbation theory[10]. The spectrum obtained by this method is nearly valley degenerate, and exactly spin degenerate. At high fields, the valley and spins are split, likely due to a combination of Zeeman effect, the band gap, and exchange effects. To simulate this, we add a phenomenological energy splitting $\delta E = (s + \xi/2)\Delta$, where $s = \pm 1$ and $\xi = \pm 1$ are spin and valley quantum numbers, respectively, and $\Delta = 8B/B_0$(meV) is the splitting width. Results of this calculation are shown in Fig. S5.

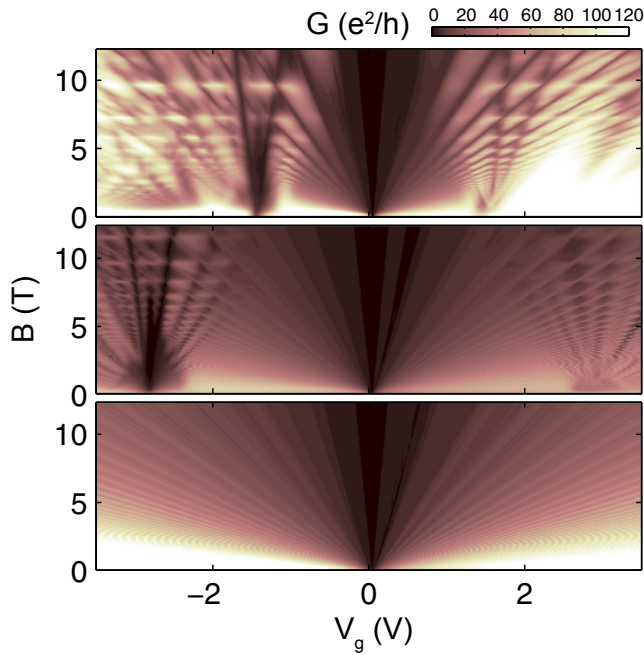

Figure S3. **Magnetotransport and superlattice Dirac points.** Top to bottom: devices A1, A2 and B2.

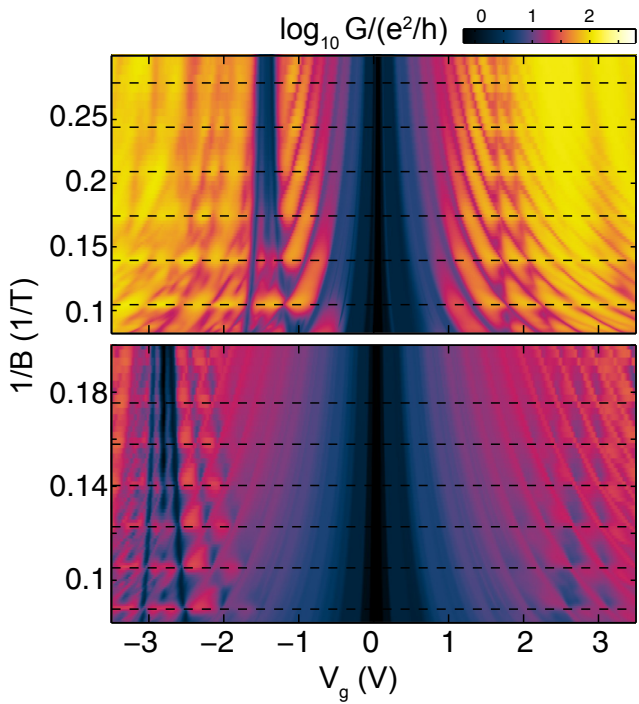

Figure S4. **Fits to Landau level crossings of Hofstadter spectrum and Hofstadter oscillations.** Magnetotransport of samples A1 (top) and A2 (bottom) plotted as a function of $1/B$. Dashed lines are a fit to the Landau level crossings with the equation $B_0/B = q$, where $q$ is an integer and $B_0$ is the flux quantum per superlattice unit cell. $B_0$ is equal to 28.7 T and 57.0 T for samples A1 and A2 respectively.

a finite moiré superlattice period $56a \approx 13.8$ nm. The in-

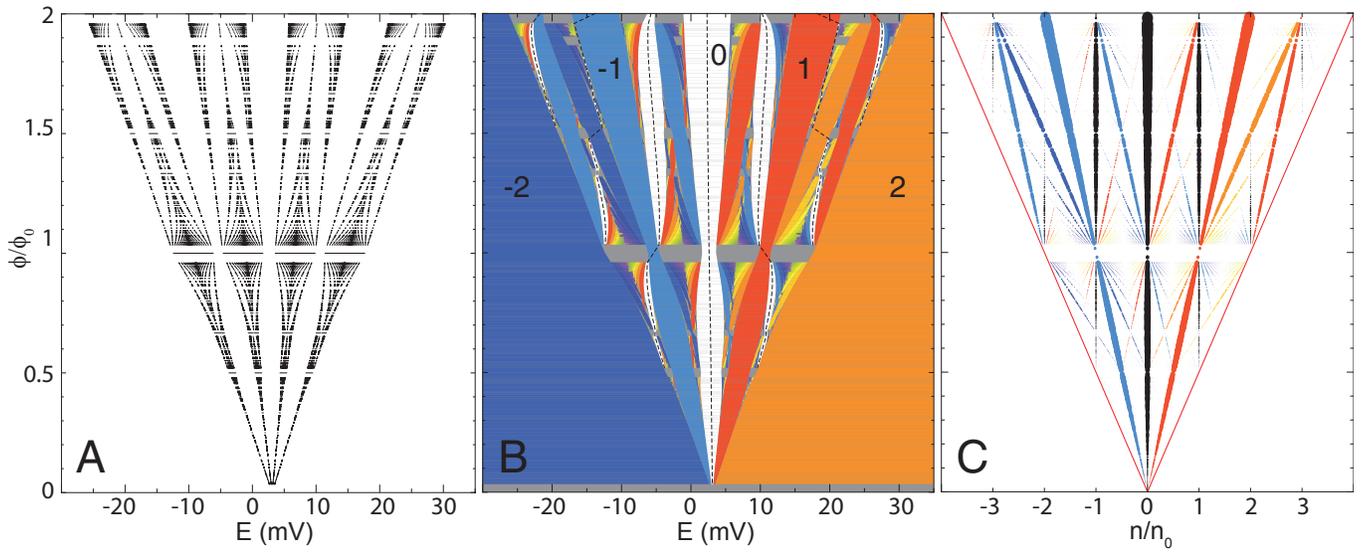

Figure S5. **Calculated Hofstadter spectrum, Hall conductivity, and Wannier diagram for the $N$=0 LL with fully broken spin and valley symmetry.** (A) Hofstadter butterfly spectrum calculated from a tight binding model with an additional valley and spin splitting. (B) The same data, with gaps color-coded to reflect the quantized Hall conductance. The numbers indicate the corresponding value of the hall conductance. The dashed curves mark constant density within the energy spectrum, specifically integer multiples of $n_0$. (C) Wannier diagram for the $N$=0 Landau level. The depicted features are gaps, color coded as in (B). Gapped features follow linear trajectories [3–5].

## IV. LANDAU LEVEL SPECTRUM

In this section we describe the single particle spectrum of monolayer graphene in the presence of a mass gap. Using the basis $(\psi_{KA}, \psi_{KB}, \psi_{K'B}, -\psi_{K'A})$, the Hamiltonian in zero magnetic field simplifies in each valley to

$$\hat{H} = \begin{pmatrix} \xi m^* v_F^2 & v_F(p_x - ip_y) \\ v_F(p_x + ip_y) & -\xi m^* v_F^2 \end{pmatrix} \quad (5)$$

where $\xi = \pm 1$ labels the valley. Working in the Landau gauge $\vec{A} = -By\hat{x}$, the Hamiltonian is independent of $x$ so that the wavefunctions are parameterized by the conserved quantum number $p_x$, where $\langle \vec{r}|\Psi\rangle = e^{ip_x x/\hbar}\vec{\phi}(y)$. After introducing the identities

$$\vec{\pi} \equiv \vec{p} - e\vec{A} \qquad \hat{a}^{(\dagger)} \equiv \frac{\ell_B}{\hbar\sqrt{2}}(\hat{\pi}_x \mp i\hat{\pi}_y) \qquad \ell_B = \sqrt{\frac{\hbar}{eB}}$$

we arrive at the Hamiltonian

$$\hat{H} = \frac{\hbar v_F \sqrt{2}}{\ell_B} \begin{pmatrix} \xi\mu & \hat{a}^\dagger \\ \hat{a} & -\xi\mu \end{pmatrix} \quad (6)$$

where $\hat{a}$ and $\hat{a}^\dagger$ obey $[a, a^\dagger] = 1$, so that they are creation and annihilation operators operating in the space of quantum harmonic oscillator states $|n\rangle$, and

$$\mu = \frac{m^* v_F^2 \ell_B}{\hbar v_F \sqrt{2}}$$

parameterizes the effective mass, written in units of the cyclotron energy. The eigenstates differ in form for $n = 0$ and $n > 0$. For the latter, there are two solutions for every positive $n$, with eigenvectors and corresponding eigenvalues

$$|\phi_{n>0}\rangle = \frac{1}{\sqrt{2}} \begin{pmatrix} \xi\sqrt{1 \pm \frac{\mu}{\sqrt{n+\mu^2}}}|n\rangle \\ \pm\sqrt{1 \mp \frac{\mu}{\sqrt{n+\mu^2}}}|n-1\rangle \end{pmatrix} \quad (7)$$

$$\epsilon_n = \pm\xi \frac{\hbar v_F \sqrt{2}}{\ell_B}\sqrt{n + \mu^2} \quad (8)$$

For $n = 0$ the solution is

$$|\phi_0\rangle = \begin{pmatrix} |0\rangle \\ 0 \end{pmatrix} \quad (9)$$

$$\epsilon_n = \xi m^* v_F^2 \quad (10)$$

The choice of basis means that the wavefunctions in the two valleys are fully polarized on different sublattices. The combined spectrum can be rewritten more transparently in terms of a quantum number $N$ spanning all integers so that we recover the spectrum given in the main text,

$$\epsilon_N = \begin{cases} \frac{\hbar v_F \sqrt{2}}{\ell_B}\text{sgn}(N)\sqrt{2(\hbar v_F)^2|N|/\ell_B + (m^* v_F^2)^2} & N \neq 0 \\ \xi m^* v_F^2 & N = 0 \end{cases} \quad (11)$$

For $m^* = 0$, the spectrum is identical for both the valleys, as shown in Figure S6A. While a nonzero $m^*$ leads to shifts in all the energy levels, it leaves the valley degeneracy of the $N \neq 0$ levels intact. Not so the $N = 0$ LL, which splits into two doubly degenerate (when real spin is accounted for) levels (Fig. S6B). Like the zero mode of the massless equation, these levels do not disperse with magnetic field (Fig. S6C).

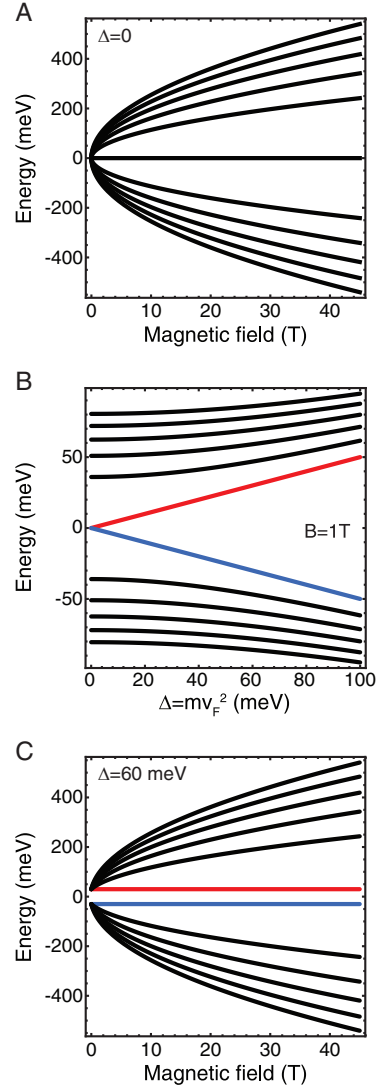

Figure S6. **Landau level energy spectrum for $-5 < N < -5$** (A) LL spectrum for $m^* = 0$ as a function of magnetic field, showing the zero mode and $\sqrt{B}$ dependence of the cyclotron energies. (B) Splitting of the zero mode with increasing $\Delta = 2m^* v_F^2$. Black curves indicate valley degenerate LLs, while red and blue indicate sublattice polarized levels in which the valley degeneracy has been lifted. Within these calculations, which neglect the Zeeman splitting, the black levels are fourfold degenerate while the red and blue levels are twofold degenerate. (C) LL spectrum with $\Delta = 60$ meV showing the sublattice polarized zero mode, which does not disperse with magnetic field.

## V. GAP MEASUREMENTS

### A. Temperature dependence of conductivity at $B=0$

The temperature dependence of the minimum conductivity of each device was measured to elucidate the nature of the observed insulating state. These data are presented in Figure S7, where the horizontal temperature axis is plotted as $1/T$. All samples exhibit a strong temperature dependence at high temperatures (200K to 20K), with the conductivity decreasing by many orders of magnitude as the temperature decreases. We fit this high temperature regime with an Arrhenius dependence $\sigma \propto \exp(-\Delta/2T)$ to extract a gap, $\Delta$, for each sample (Fig. S8). An Arrhenius dependence describes the high temperature conductance variation, especially in the samples with the largest $\Delta$ (samples A1 and A2). As the temperature is further reduced the conductivity deviates from a single gap Arrhenius law (Fig. S7), with samples B1 and B2 showing the largest deviations. At low temperatures ($T < 20K$), the temperature dependence is much weaker, possibly indicating the onset of hopping-dominated transport [11]. Additional fluctuations in the conductance vs gate measurements are also evident in this regime, which we ascribe to mesoscopic effects.

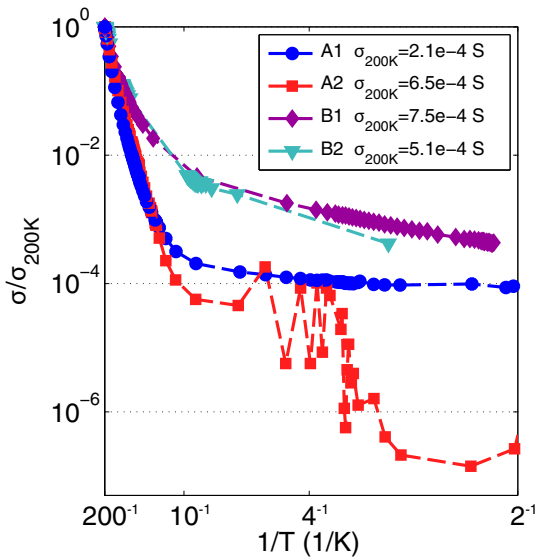

Figure S7. **Temperature dependence of conductivity from 200K to 2K.** Normalized conductivity plotted against $1/T$. All samples exhibit a strong temperature dependence at high temperatures (200K to 20K) and a weaker dependence for $T < 20K$.

### B. Magnetic field dependence of the gap

$\Delta$ is not field independent at low fields, varying non-monotonically as the field is increased (Fig. S9). The gap is determined by fitting the temperature dependence of the minimum conductivity for each sample to an Arrhenius dependence from $T = 20\text{-}50K$ over a factor of 10 change in con-ductance at various magnetic fields. We note that this effect is absent in the single-particle model of [12].

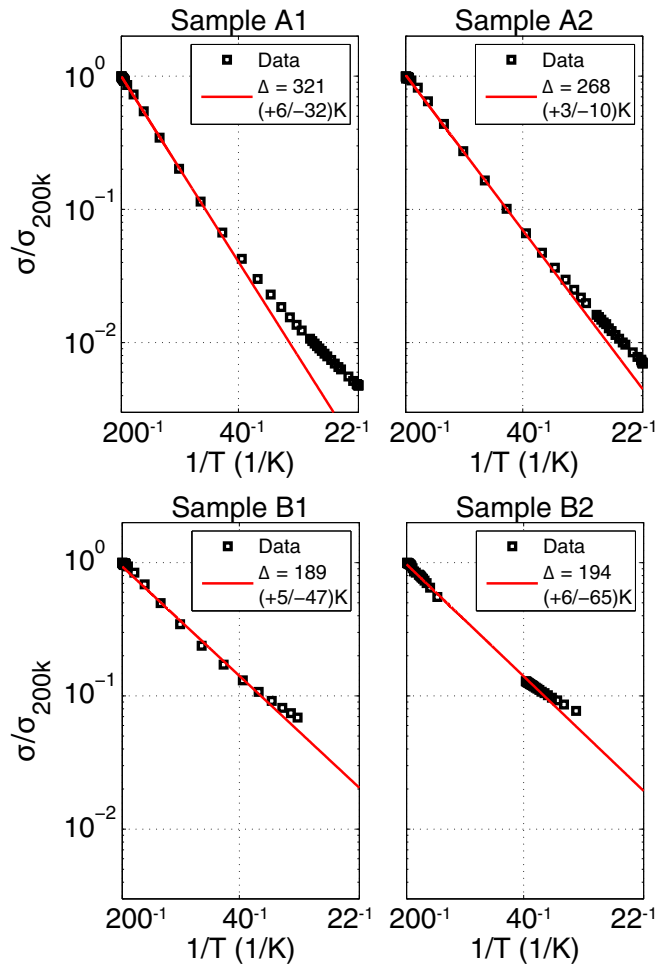

Figure S8. **Arrhenius fits.** Data (black symbols) and high temperature fits (red line). Measured values of $\Delta$ are presented in the legend for each sample along with the spread in gap values (in parentheses) due to uncertainty in the range of the high temperature regime. These gap values appear in Figure 4C of the main text.

## VI. CAPACITANCE MEASUREMENTS

### A. Principle of measurement

We measure capacitance between the graphene sheet and the graphite back gate by using a low-temperature capacitance bridge, based on a high-electron mobility transistor (HEMT). Our "bridge-on-a-chip" has very high sensitivity (approximately $10^{-1} e/\sqrt{\text{Hz}}$; $e$ is the electron charge) and has been used to measure, among other things, single-electron charging of semiconductor quantum dots [13]. This technique is particularly well suited for measuring small-area graphene devices. It is based on the principle of applying a known AC excitation $V_{meas}$ to the top plate of the unknown capacitance $C_{meas}$ and a second AC excitation $V_{std}$, approximately 180° out of phase with $V_{meas}$ to the top plate of a known, standard capacitor $C_{std}$ Figure S10. The bottom plates are connected at a common point–the "balance point"– and $V_{std}$ is adjusted until the potential modulation at the balance point is zero. The unknown capacitance is then given by

$$C_{meas} = (V_{std}/V_{meas})\, C_{std}. \qquad (12)$$

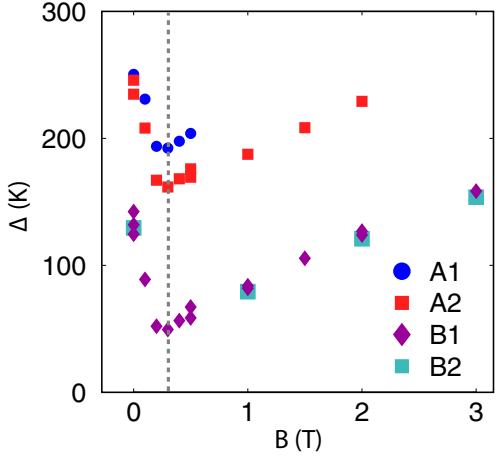

Figure S9. **Dependence of Arrhenius gap on magnetic field.** Gap extracted by fitting to Arrhenius behavior from 10 K to 50 K at different magnetic fields. Gaps for all samples exhibit nonmonotonic behavior, with a minimum at $B \simeq 0.25$ T (gray dashed line).

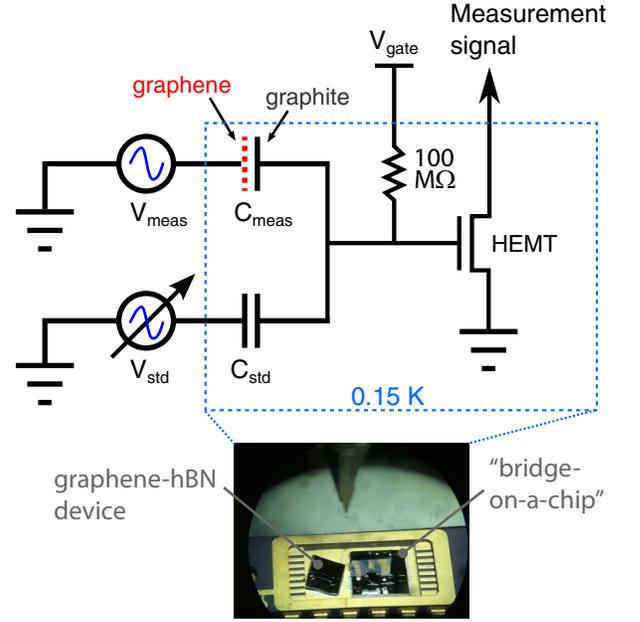

Figure S10. **Schematic of high-sensitivity capacitance measurement.**

We typically use an excitation $V_{meas}$ of 100 $\mu$V to 1 mV.

In practice, to measure the capacitance $C_{meas}(V_g, B)$, we balance once at the beginning of a $V_g$ sweep and then measure the off-balance signal to determine the capacitance. Placing the gate of the HEMT at the balance point reduces the capacitance to ground at that point by a factor of $C_{gs}/C_{cable}$, where $C_{gs} \approx 0.4$ pF is the gate-source capacitance of the HEMT and $C_{cable} \approx 200$ pF is the capacitance to ground of the coaxial cable used to measure the signal at the (cold) balance point. Thus, even when the gain of the transistor is of order unity (and in practice we operate at higher gain of $\sim 3-4$), the transistor mitigates the shunting effect of the coaxial cable and consequently improves the signal-to-noise by a factor of $C_{cable}/C_{gs} \approx 500$.

### B. Additional capacitance measurements

When the bulk resistance $R$ of the device becomes large, the equation relating the measured capacitance $C_{meas} \equiv C$ and the density of states,

$$\frac{1}{C} = \frac{1}{C_{geom}} + \frac{1}{eA(\partial n/\partial \mu)}, \qquad (13)$$

is no longer valid when the measurement frequency $\omega$ is larger than $\sim (RC)^{-1}$. This can happen, for example, when the chemical potential $\mu$ lies in the gap between Landau levels. Measuring a reduction in capacitance in this regime reflects the inability of the device to charge on the time scale $\omega^{-1}$, rather than a direct measurement of the density of states as per Equation 13 [14]. However, dips in the capacitance still qualitatively signify the formation of a gap in the energy spectrum.

The resistance of the device will introduce a phase shift in the balance-point signal. We keep track of this phase shift by recording both the $X$ (capacitive) and $Y$ (resistive) quadratures of the balance-point voltage. In finite field, as the gate voltage (and chemical potential) are swept, the measurement oscillates between the high-frequency regime $\omega > (RC)^{-1}$, when $\mu$ lies in a Landau gap, and the low-frequency regime $\omega < (RC)^{-1}$, when filling a highly-degenerate Landau level. Transitions between the two regimes are marked by pronounced peaks in the $Y$ quadrature, as can be seen in Figure S11.

For measurements at higher frequencies, a small phase shift $\delta$, unrelated to the resistance of the device, can be introduced into the measurement by, for example, mismatched cables or attenuators. We rotate the signal according to $X' = X\cos(\delta) - Y\sin(\delta)$ and $Y' = X\sin(\delta) + Y\cos(\delta)$, where $\delta$ is chosen such that $Y' \equiv 0$ in a highly compressible regime where $C \approx C_g$, e.g when $\mu$ lies in an orbital Landau level at low field. Figure S12 shows capacitance $C = X' \cdot C_{std} - C_p$ and "loss" (defined as $Y' \cdot C_{std}$) for the semimetallic and insulating graphene capacitors (shown in Fig. 3 main text). Here $C_p$ is the parasitic background capacitance due to, for example, the capacitance between adjacent wire bonds or pins on the device mount.

Capacitance measurements are sensitive to the bulk of the sample in both the high and low frequency limits, making it a complementary measurement technique to transport. For example, Fig. S13 shows both transport and capacitance data taken on device A1 in the regime of Hofstadter minigaps. Well developed quantum Hall plateaus in conductance have corresponding dips in capacitance indicating a poorly conducting bulk associated with a spectral gap. Remarkably, dips in capacitance appear even when the expected transport features associated with Hofstadter minigaps are poorly developed.

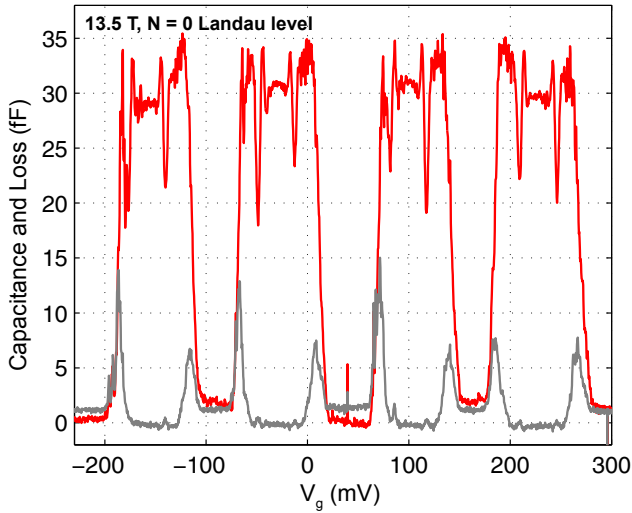

Figure S11. **Capacitance and loss of the insulating graphene capacitor in the $N = 0$ Landau level at 13.5T** Capacitance is shown in red and loss in grey.

### C. Extracting an upper bound on the gap $\Delta\mu$ from gate dependence of transport and capacitance

In Figure 4C of the main text, we present an additional estimate of the bandgap by taking the width in gate voltage $\Delta V$ for the region defined by $\sigma < 0.1e^2/h$. This estimate is based on the assumption that conductance exponentially increases when the gate voltage is large enough to overcome the gap at the CNP. The chemical potential change ($\Delta\mu$) across this insulating region can be estimated from capacitance measurements, which, in the low frequency limit, can be analyzed to extract

$$\Delta\mu = \int_{dip} e\left(1 - \frac{C}{C_{geom}}\right) dV_g, \qquad (14)$$

which follows from Eq. 13 after substituting $dn = (eA)^{-1} C dV_g$. Fig. S14 shows transport and capacitance data from device B2. Over the range defined by $\sigma < 0.1e^2/h$, the capacitance is never greater than 20% of the geometric capacitance. Compared with our naive estimate, then, the second term under the integral reduces the $\Delta\mu$ by $\approx 20\%$. We thus can apply this estimate even when capacitance data is not available, producing an upper bound on $\Delta\mu$. This estimate is enabled by the high geometric capacitance of our devices, $C_g$, which suppresses disorder-induced localized state contributions $\Delta\mu$.

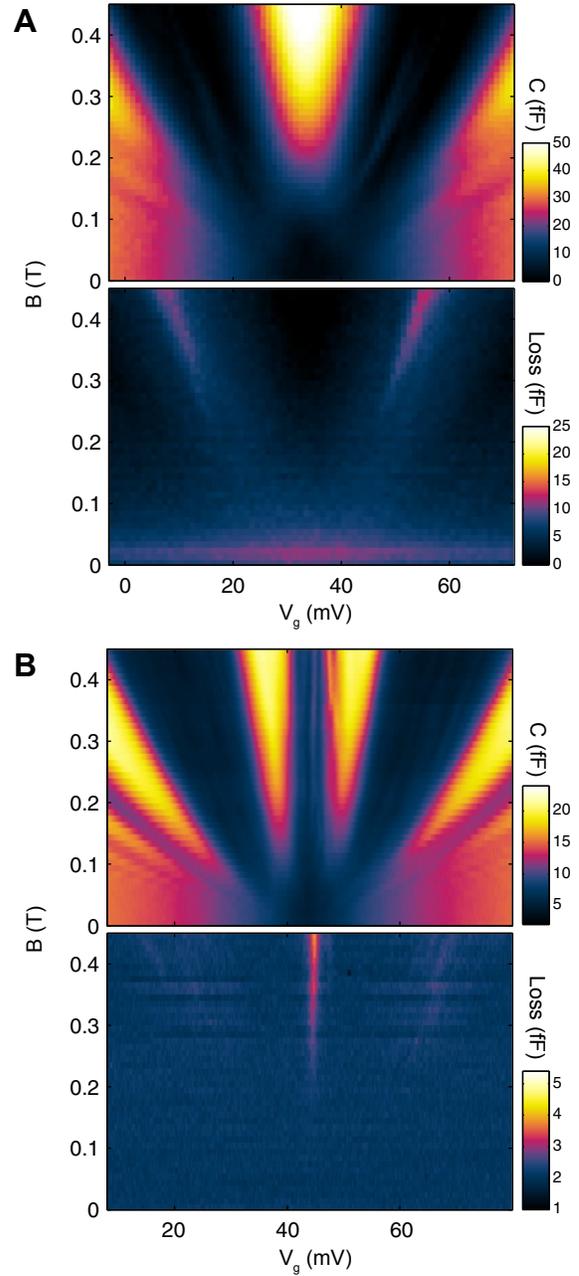

Figure S12. **Capacitance and loss of the semimetallic and insulating graphene-hBN capacitors.** (**A**) The same measurement of the capacitance of the semimetallic graphene capacitor depicted in Fig. 3A of the main text, alongside a concurrent measurement of the loss signal. The measurement frequency was 173.5 kHz. A parasitic background capacitance $C_p$=2 pF was subtracted from the capacitance signal after a rotation of $\delta = -7.2°$. The scale of the capacitance color bar is twice as large as that of the loss. (**B**) The same measurement of the capacitance of the insulating graphene capacitor (device B1) depicted in Fig. 3B, alongside a concurrent measurement of the loss signal. The measurement frequency was 56.2 kHz. A parasitic background capacitance $C_p$=57 fF was subtracted from the capacitance signal after a rotation of $\delta = -1.1°$. The scale of the capacitance color bar is five times as large as that of the loss.

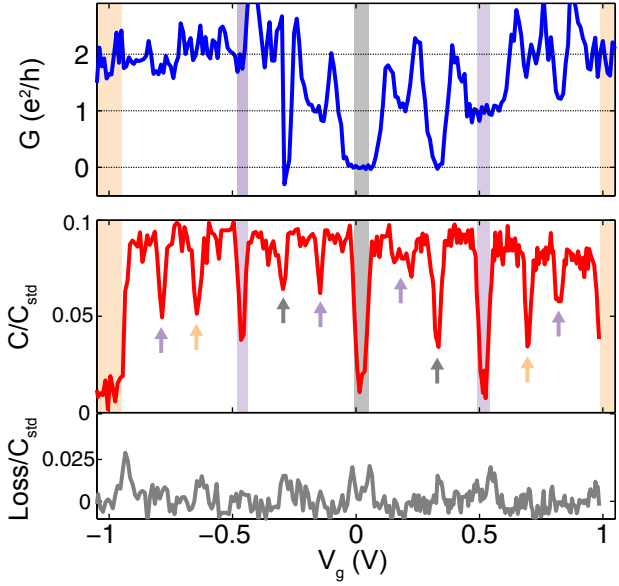

Figure S13. **Comparison of transport and capacitance at 45 T.** Top: transport measurement. Features are similar to the measurement in Fig. 2C, top panel (at 43T). Bottom: capacitance and loss measurement at 45 T. Measurement frequency was 48 kHz. Shaded rectangles indicate normal quantum Hall states at filling factors $\nu = 0, \pm1$ and $\pm2$, as in Fig. 2C. Colored arrows indicate bulk insulating states associated with Hofstadter minigaps in the $N = 0$ LL for $\phi/\phi_0 > 1$.

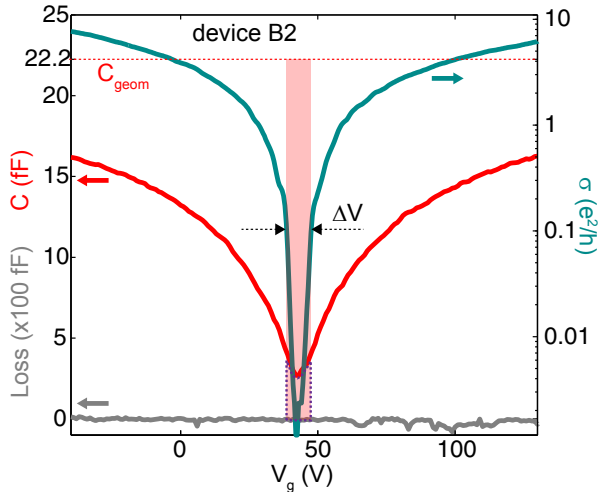

Figure S14. **Estimate of zero-field gap from chemical potential change $\Delta\mu(\Delta V)$.** Simultaneous measurement of zero-field capacitance (red) and conductivity (blue) of device B2. Over a range of gate voltage $\Delta V$, $\Delta\mu = e\Delta V - \int (C/C_{geom})dV$ (Equation 14); that is, the shaded red rectangle minus the area indicated by the purple outline. Here, $C_{geom}$=22.2 fF is indicated by a dashed red line. Thus, $\Delta\mu = e\Delta V$ serves as an upper bound on the magnitude of the gap.

## VII. SAMPLE FABRICATION

### A. Sample Design

Figure S15 contains a schematic of the graphene-boron nitride heterostructures, which were fabricated by sequential stacking of graphite, hBN, and graphene onto resistive silicon wafers with 295 nm of thermally-grown oxide (resistance 10 k$\Omega$ per square, from Nova wafer). The wafer was chosen to provide ideal optical contrast for thin hBN and graphite flakes, but with insulating silicon to avoid parasitic capacitances associated with a metallic back plane at low temperature. The bottom graphite layer provides a local gate for controlling the graphene carrier density as well as for performing capacitance measurements. The graphite also performs a key role as an ideal substrate layer for the graphene-hBN heterostructure by screening charge disorder from the underlying SiO$_2$ substrate and also providing additional screening of charge inhomogeneity in the graphene [15]. We believe that the usage of a local graphite substrate with a thin hBN dielectric layer is vital to observing the intrinsic behavior of the graphene-hBN heterostructure. Graphite is an ideal substrate material since it bonds well to both hBN and graphene, providing an atomically-flat interface which is resistant to breakdown or reconstruction during the various fabrication steps.

### B. Specific Geometry of Presented Devices

All four of the devices discussed in the main text originate from the same graphene flake transferred onto a single hBN flake. A topographic AFM image of the original graphene flake after transfer onto the hBN is shown in Figure S15E. After transfer, the graphene flake has a few isolated wrinkles (size $\sim$1-80nm) which separate micron-sized regions of ultra-flat graphene. The four devices are etched from the flat regions of the graphene, with each device being separated from the other by a graphene wrinkle. In general, it can be expected that such ripples will introduce small angular rotations between each of the isolated pieces, despite having originated from the same graphene flake.

### C. Nanofabrication Process

The steps taken to produce the devices discussed in the main text are as follows:

1. A Si wafer was hand-cleaved by a diamond scriber and then treated by a mild O$_2$ plasma for 2 mins to remove surface contaminants and particles. Graphite (Natural Graphite, NGS Naturgraphit) was then exfoliated using Ultron systems R1007 tape and immediately soaked in acetone and IPA to remove tape residue and to dislodge loosely attached flakes.

2. Thin Graphite flakes were identified (5-25nm thick) and etched into rectangular bars (10umx100um) using O$_2$ Plasma and a PMMA etch mask (950 A5 @ 2000rpm). The pmma mask was removed in acetone and IPA and the chip+graphite was heat cleaned in a quartz tube furnace at 350C for 5 hours under Ar:H2 flow. Each flake was then imaged in an atomic force microscope (AFM) to ensure that they were atomically flat and clean, with no resist residue or other particulate contaminants on the surface. The most common problem at this step is overetching, which crosslinks the PMMA mask, resulting in nm size particles on the graphite which cannot be removed by heat cleaning.

3. Thin boron nitride flakes were transferred onto the graphite bars using the dry transfer method outlined in the next section. The samples were then heat cleaned and AFM imaged to check for regions that are flat and clean. Typically there are many bubbles trapped between the graphite and boron-nitride layers, with occasional flat regions that are many microns in size.

4. Graphene flakes were transferred onto the flat regions of the h-BN + graphite stack. The samples were then heat cleaned and AFM imaged to check for flat regions without bubbles or wrinkles. Often the graphene flakes would appear not completely clean even after heat cleaning, but this was disregarded at this step since additional sample processing and heat cleaning generally increases the cleanliness of the sample.

5. The samples were then contacted, using ebeam lithography to design a PMMA mask (bilayer resist with 495 A4 @ 2000rpm and 950A2 @ 3000rpm, cold developed in H2O:IPA 1:3 to avoid pmma cracks on the hBN) and then evaporated with 1nm Cr:80nm Au in a thermal evaporator (Cr layer is kept thin to avoid strain damage during heating/cooling of device). Then, the graphene flakes were etched into the desired geometry, isolating regions that were flat and free of bubbles and wrinkles by using a PMMA etch mask and RIE O$_2$ plasma etching. The final step was then to heat clean for 6 hours under Ar:H2 flow at 350C. Often graphene flakes which appeared dirty after the initial heat clean step would become very clean after the final heat cleaning step.

### D. Flake Dry Transfer Method

To fabricate the graphene-hBN heterostructures we use a dry transfer method inspired by [16]. The technique centers on a stack of transparent materials which support a polymer release layer onto which graphene and hBN flakes are exfoliated. This polymer release layer with graphene and hBN can then be aligned and brought into contact with a target substrate at high temperature, which causes the release layer to adhere to the target and detach from its support.

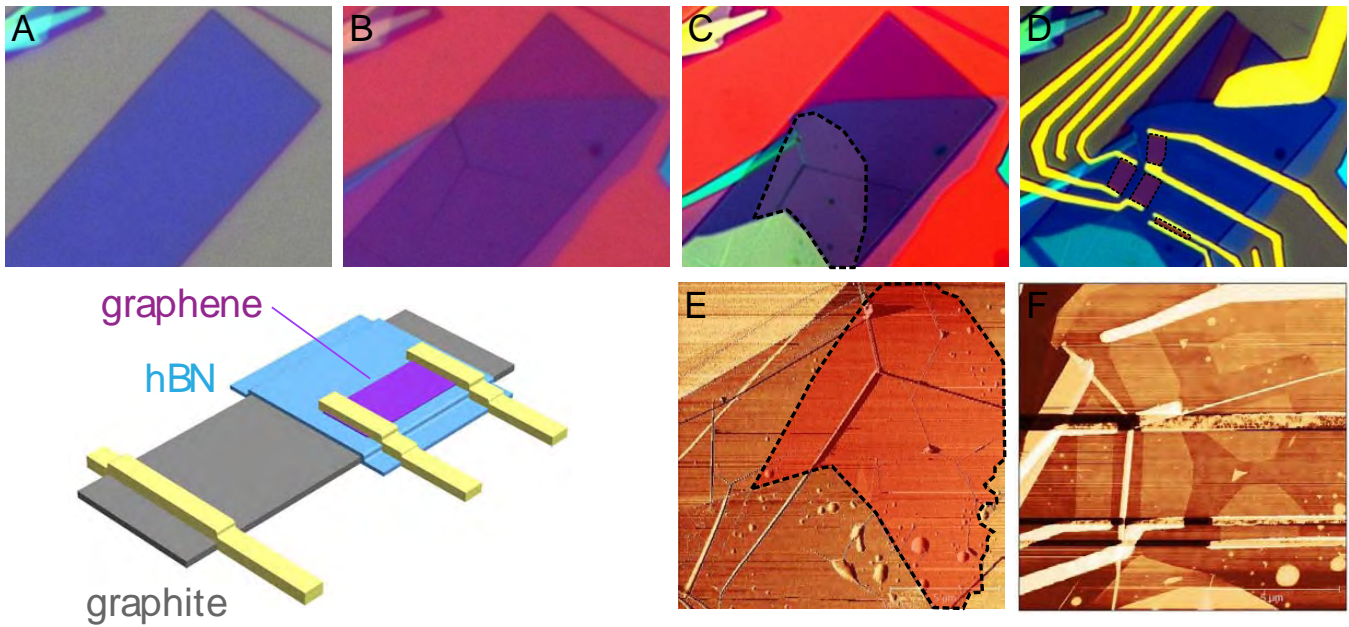

Figure S15. **Overview of device fabrication steps and device schematic.** A-D are optical images and E/F are AFM images of the fabrication steps for the devices discussed in the main text. Scale of optical images is 32.5um wide **(A)** Etched graphite bar on Si/SiO$_2$ wafer. **(B)** After transfer of 7 nm-thick hBN. **(C)** After graphene transfer (dash line depicts graphene boundary). **(D)** Final contacted and etched device (graphene strips are false colored red). **(E)** AFM image of graphene (red false color) on hBN. **(F)** AFM image of final contacted and etched device .

The transparent support stack consists of a glass slide, a layer of 1mm thick PDMS (Sylgard 184) and a layer of Duck HD clear packing tape. The PDMS is exposed in a UV ozonator for 15 mins before placing the tape layer to increase adhesion. Next, two layers of MMA (8.5 EL, Microchem) are spun at 1200rpm for 70s onto the top tape layer, with a 10min bake step at 180C after each spin. The MMA layer is the polymer release layer which supports and transfers the flake materials. The tape's purpose is to increase the adhesion of the MMA to the PDMS, and also blocks contamination from the PDMS. The PDMS is a flexible support layer that can deform to ensure conformal contact between the polymer release layer and the target substrate. The glass slide is a sturdy transparent support to make handling easier. All layers are transparent such that a target substrate can be seen through the support layer during the transfer process.

Graphene and hBN flakes are exfoliated onto the transparent support stack using wafer backing tape. Desired flakes are then scanned for in an optical microscope in reflection mode and monolayer graphene flakes can be easily found. The location of desired flakes are marked on the bottom of the glass slide with a marker and then cut out using a sharp scalpel into a 3mmx3mm square. This square PDMS+tape+MMA+graphene/hBN piece is then removed with clean tweezers and affixed to the end of a glass slide. This glass slide is mounted into a home built transfer alignment system, consisting of a micromanipulator under a high working distance microscope (Bausch&Lomb Microzoom) with a heated stage. The target substrate is mounted onto the heated stage, and then is aligned under the transfer slide with the MMA release layer + graphene/hBN facing towards the target. The two are then brought into careful contact while adjusting the alignment with the stage heated to 35°C. Once in contact the stage is heated to 130C, and then the transfer slide is disengaged from the target. At this point the MMA will detach from the tape and the transparent support and remain stuck to the target SiO$_2$ substrate.